# Signal Acquisition of Luojia-1A Low Earth Orbit Navigation Augmentation System with Software Defined Receiver

**Liang Chen[1], Xiangchen Lu[1], Nan Shen[1,*], Lei Wang[1], Yuan Zhuang[1], Ye Su[1], Deren Li[1] and Ruizhi Chen[1]**

**Abstract** Low earth orbit (LEO) satellite navigation signal can be used as an opportunity signal in case of a Global navigation satellite system (GNSS) outage, or as an enhancement means of traditional GNSS positioning algorithms. No matter which service mode is used, signal acquisition is the prerequisite of providing enhanced LEO navigation service. Compared with the medium orbit satellite, the transit time of the LEO satellite is shorter. Thus, it is of great significance to expand the successful acquisition time range of the LEO signal. Previous studies on LEO signal acquisition are based on simulation data. However, signal acquisition research based on real data is very important. In this work, the signal characteristics of LEO satellite: power space density in free space and the Doppler shift of LEO satellite are individually studied. The unified symbol definitions of several integration algorithms based on the parallel search signal acquisition algorithm are given. To verify these algorithms for LEO signal acquisition, a software-defined receiver (SDR) is developed. The performance of those integration algorithms on expanding the successful acquisition time range is verified by the real data collected from the Luojia-1A satellite. The experimental results show that the integration strategy can expand the successful acquisition time range, and it will not expand indefinitely with the integration duration.

[1] State Key Laboratory of Information Engineering in Surveying, Mapping and Remote Sensing, Wuhan University, Wuhan 430079, China

Liang Chen l.chen@whu.edu.cn

Xiangchen Lu 2018286190164@whu.edu.cn

✉ Nan Shen nanshen@whu.edu.cn

Lei Wang lei.wang@whu.edu.cn

Yuan Zhuang yuan.zhuang@whu.edu.cn

Ye Su sueyeah@whu.edu.cn

Deren Li drli@whu.edu.cn

Ruizhi Chen ruizhi.chen@whu.edu.cn

## Introduction

Global navigation satellite system (GNSS) has been widely used in navigation, positioning, timing, precision agriculture [1], structural health monitoring (SHM) [2], [3], remote sensing [4],[5], and other fields. However, GNSS positioning is affected by various measurement errors, such as ionospheric delay, tropospheric delay, and multipath effect. Besides, the GNSS signal is very weak, it comes from 20,000-30,000 kilometers away and is vulnerable to unintentional radio frequency interference or malicious interference (jamming and spoofing) [6], [7]. Therefore, it is of great significance to enhance the reliability and positioning accuracy of GNSS by other means [8]. Many studies consider using signals of opportunity (SOP) for positioning when GNSS is unavailable or unreliable. These signals of opportunity include digital television [9]-[11], Bluetooth [12], low earth orbit [13], [14], WIFI [15]-[17], vision[18],[19], and 5G [20]-[22], etc. Among them, the LEO satellite has been paid more and more attention and has become a research hotspot.

On the one hand, LEO is studied as a non-GNSS alternate for positioning in case of a GNSS outage. In [13], the performance of doppler positioning using one LEO satellite has been analyzed. The results showed that doppler positioning based on full one pass data can achieve an accuracy that less than 100 m most of the time. A framework to navigate with of LEO satellite signal was proposed, of which pseudo-range and Doppler measurements of the LEO satellite were used to aid inertial navigation [14]. Simulations were carried out in different scenarios, including GNSS partially or completely unavailable, different numbers of LEO, and the position of LEO known or unknown.

On the other hand, LEO is being studied as an enhancement means of traditional GNSS positioning algorithms. In [23], a study on accelerating precise point positioning (PPP) convergence time by combining global positioning system (GPS) and LEO was carried out. The simulation results show that compared with the GPS, the PPP convergence time of GPS/LEO is reduced by 51.3%, and the accuracy is also improved by 14.9%. In [24], an LEO-augmented full operational capability (FOC) multi-GNSS algorithm for rapid PPP convergence was proposed. Different LEO constellations were designed and complicated simulations were performed. The results show that the convergence time of PPP is significantly reduced as the number of visible LEOs increases. In the meanwhile, the rapid motion of LEO satellites also contributes to geometric diversity and enables rapid convergence of PPP. The LEO enhanced GNSS (LeGNSS) system concept was proposed to improve the performance of the current multi-GNSS real-time positioning service in [16], where different operation modes and schemes of the LeGNSS system are introduced and analyzed.

Regardless of the LEO service mode mentioned above, signal acquisition is a prerequisite for providing enhanced low-orbit navigation services. LEO satellite orbit is different from that of the GNSS satellite, which results in different Doppler frequency shifts [25]. Besides, there are few studies on LEO navigation augmentation signal acquisition, and simulation data are mostly used even if they exist [26], [27]. There are many research results about the acquisition algorithms of the GNSS signal, and it is necessary to verify the applicability of these acquisition algorithms for the LEO navigation augmentation signal. Compared with the medium orbit satellite, the transit time of the LEO satellite is shorter, so it is of great significance to expand the successful acquisition time range of the navigation augmentation signal.

The Luojia-1A satellite is a lightweight scientific LEO satellite designed by Wuhan University for night light remote sensing [28] and LEO signal navigation augmentation experiments [25], [29], which is based on the concept of integrated communication, navigation, and remote sensing [30]-[32]. The satellite was launched from Jiuquan Satellite Launch Center in China on June 2, 2018, with an orbit height of 645 kilometers. The satellite is equipped with three L-band antennas, two of which are used to receive GPS/Beidou signals and one is used to broadcast navigation augmentation signals. The software-defined receiver is adopted to study LEO navigation augmentation signal acquisition in this research for several reasons. Firstly, the transit time of the Luojia-1A satellite is very short, so it is very important to collect data first and analyze them afterward. Secondly, the acquisition algorithm can be tested freely by the software-defined receiver, which has great flexibility. Besides, for the algorithm verification of this experimental satellite, the hardware implementation of the algorithm is expensive and time-consuming.

The purpose of this paper is to explore different acquisition algorithms for the navigation augmentation signal of the Luojia-1A satellite and try to expand the available time range of the LEO signal by an appropriate acquisition algorithm. Firstly, the signal model of the Luojia-1A satellite is given, and the power spatial density and Doppler frequency shift at the ground station are analyzed. Secondly, the parallel code phase search acquisition algorithm is introduced, and several integration algorithms for weak signal acquisition are described. Thirdly, the experiment and results are presented. Then, thresholds of detection indicators and the relationship between integration duration and successful acquisition time are discussed. Finally, the conclusions are given in the last section.

## LUOJIA-1A SATELLITE SIGNAL MODEL AND CHARACTERISTICS

To study the acquisition algorithm of the Luojia-1A satellite navigation augmentation signal, the signal model is given at first. To study the signal characteristics of the Luojia-1A satellite, the Doppler frequency shift and power spatial density of GPS and the LEO satellite is compared.

### Signal Model

As an enhanced navigation satellite, the navigation signal of the Luojia-1A satellite is similar to that of the GNSS satellite [33]. The transmitted navigation augmentation signal contains three parts: carrier, navigation data, and spreading sequence. These signals are modulated onto the carrier signal by using the binary phase-shift keying (BPSK) method. Besides, the navigation data of the Luojia-1A satellite is transmitted at a rate of 50 bps. This results in a possible data bit transition every 20 milliseconds (ms), which should be considered in signal acquisition. The navigation augmentation signal emitted by the Luojia-1A satellite can be described as:

$$s(t) = A_c C(t) D(t) \, sin( 2\pi f_{H_1} t) \tag{1}$$

where $A_c$ is the amplitude of coarse/acquisition (C/A) code; $t$ denotes the time; $C(t)$ is the spreading sequence of C/A code; $D(t)$ is the navigation data; $f_{H_1}$ is the frequency of the carrier $H_1$. Normally, the input signal needs to be down-converted to a low-frequency signal for processing. The low-frequency component of down-conversion is called the intermediate frequency (IF) [34]. Downconversion is the frequency shifting in the spectrum that can be achieved by mixing the input signal with a locally generated signal [1]. The down-converted form of this navigation augmentation signal can be described as:

$$s(t) = A_c C(t) D(t) \sin\left(2\pi f_{IF} t\right) \tag{2}$$

where $f_{IF}$ is the intermediate frequency (IF). After analog-to-digital conversion, the signal can be described as:

$$S_{IF}(n) = C(n)D(n)\sin\left(2\pi f_{IF} n\right) + e(n) \tag{3}$$

where $n$ is the discrete sample point, $e(n)$ is the additive band-limited white Gaussian noise (AWGN).

To demodulate the information in the signal, the Doppler shift and code delay of the signal must be accurately obtained. The coarse Doppler shift and code delay are obtained by signal acquisition, and these parameters are passed to the tracking module to accurately obtain the Doppler shift and the code delay for signal demodulation. Therefore, signal acquisition plays an important role in the entire signal processing process. However, due to the difference between the orbits of LEO satellite and GPS satellite, as well as the different system designs, different factors need to be taken into consideration when performing LEO navigation augmentation signal acquisition.

### Large Variation of Distance and Signal Strength

Due to the Luojia-1A satellite orbit is close to the earth, as well as the dramatically varied distance between the user and the satellite, there is a large signal strength variation [25]. For a user on the earth, the shortest visible distance from the user to the Luojia-1A satellite is about 650 km, and the farthest visible distance can reach 2000 km. Besides, the variation from the most recent visible distance to the farthest visible distance occurs in about five minutes. The GPS satellite transmitting antenna is designed to set different gains for different directions according to the power loss of different propagation distances of the signal [35]. The dramatically varied distance is considered in the design of the Luojia-1A satellite transmitting antenna. Also, unlike the GNSS satellite, navigation augmentation is usually only one of the tasks of the low-orbit satellite. Luojia-1A satellite is designed to provide positioning, navigation, and timing (PNT) services and remote sensing service, as well as communication service [25]. This is the so-called ’PNTRC’ concept [36]. Therefore, other mission requirements of the satellite are probably taken into consideration in the design of antenna gain. It can be seen from the following experiments that as the distance from the user to the satellite increases, the signal strength decreases. The distance from the ground station to the GNSS satellite as well as the Luojia-1A satellite and the corresponding power space density calculated from the distance are presented in Fig. 1.

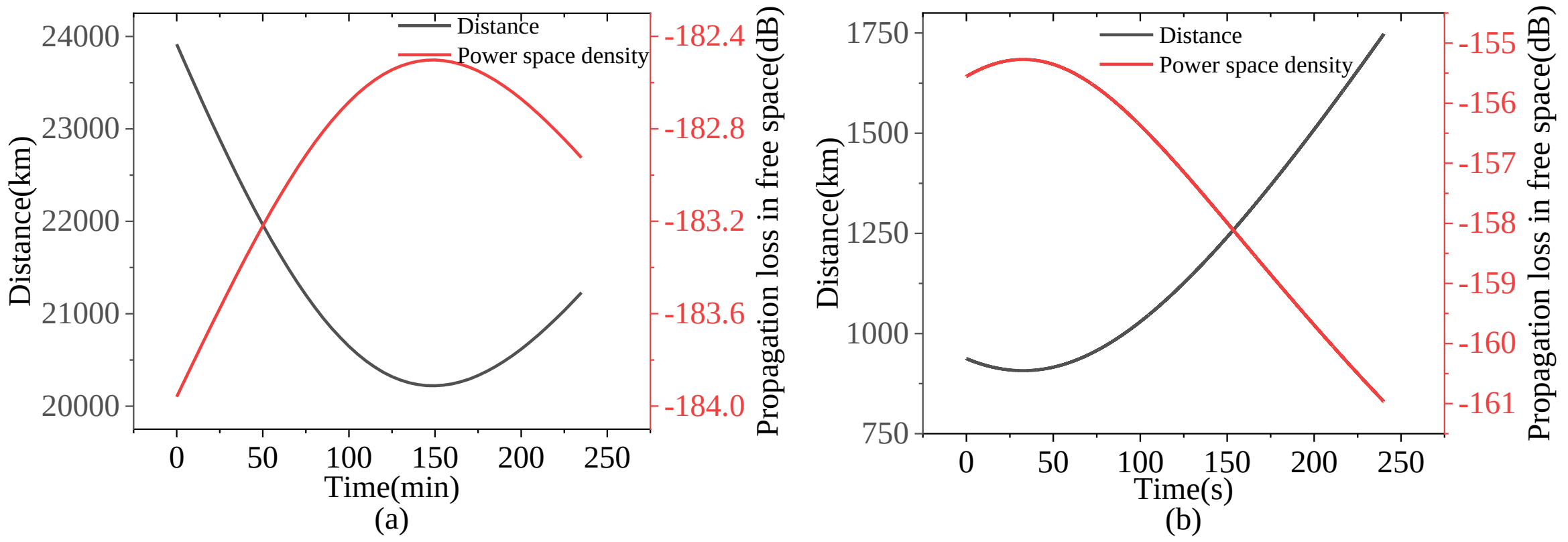

**Fig. 1**. Distance from the ground station to GNSS satellite as well as the Luojia-1A satellite and the corresponding propagation loss in free space calculated from the distance. (a)GPS ;( b) Luojia-1A.

As can be seen from Fig. 1, the distance from the ground station to the GNSS satellite and the distance to the Luojia-1A satellite are not of the same level. The large variation of the distance from the ground station to the Luojia-1A results in a larger range of free space propagation loss than that of GNSS which should be taken into consideration when designing signal acquisition algorithms.

Large Doppler Frequency Shift Range and High Doppler Frequency Shift Rate

In satellite navigation and positioning, the Doppler effect is caused by the relative radial motion between the satellite and the receiver. Due to the Doppler effect, the frequency of the received carrier signal changes, limiting the length of the data used to capture the signal, increasing the complexity of signal acquisition. The frequency variation due to the Doppler effect is called the Doppler frequency shift. The Doppler frequency shift can be expressed by the following formula [34]:

$$f_d = \frac{f v_d}{c} \tag{4}$$

where $f_d$ is the Doppler frequency shift; $f$ is the carrier frequency; $v_d$ is the relative radial speed between the receiver and the satellite; $c$ is the speed of light in vacuum. For GPS satellites, if the receiver is in low-speed motion, the Doppler frequency shift is about 5 kHz; if the receiver is in high-speed motion, the Doppler frequency shift is about 10 kHz [34] [37]. In contrast to the GPS satellite, due to the fast geometry change of the Luojia-1A satellite, there is a large Doppler variation, which affects the signal acquisition efficiency [25].

For the ground station, the radial velocity can be estimated by radial distance variation, and the formula is expressed as follows:

$$v_d = \frac{dr}{dt} \tag{5}$$

where $dr$ is the distance change from the ground station to the satellite during the time interval $dt$. The Doppler frequency shift and the Doppler frequency shift rate of the ground stationary station relative to the GPS satellite and the Luojia-1A satellite are calculated according to (4) and (5), respectively. The result is shown in Fig. 2.

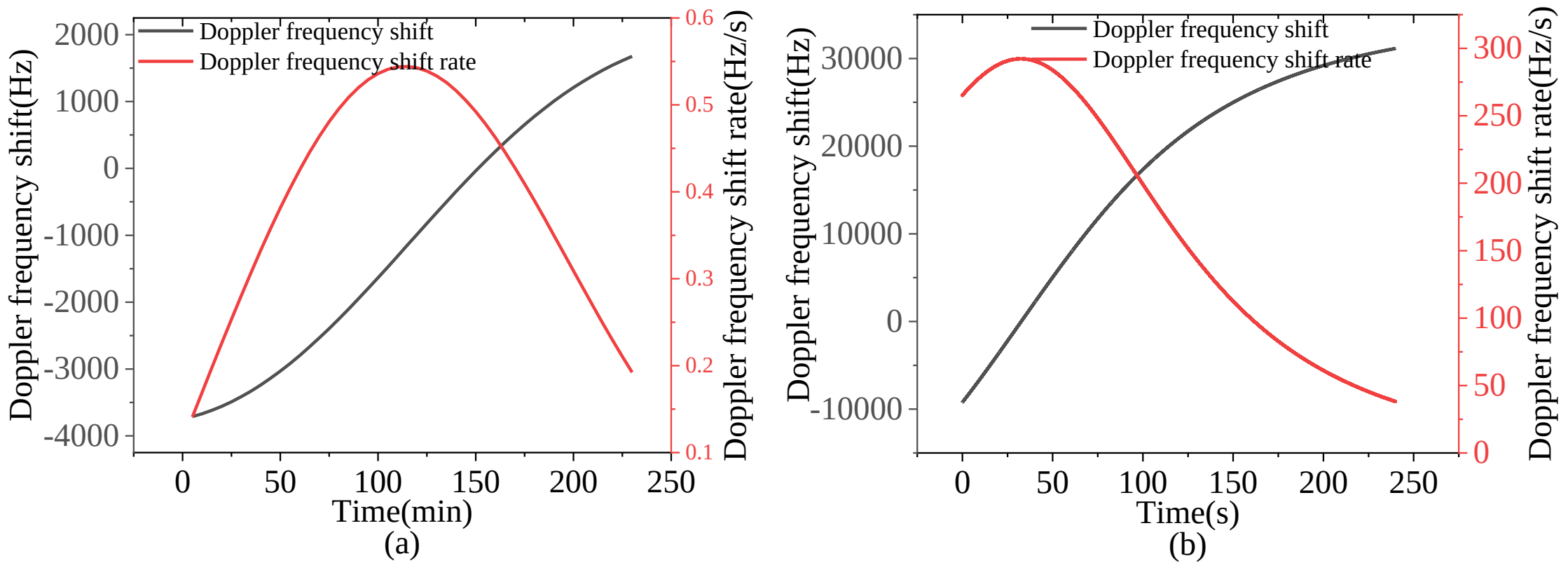

**Fig. 2.** Doppler frequency shift and the Doppler frequency shift rate of the ground stationary station relative to the GPS satellite and the Luojia-1A satellite. (a) GPS; (b) Luojia-1A.

It can be seen from Fig. 2 that the Doppler frequency shift and the Doppler frequency shift rate of the low-orbit satellite are much larger than those of the medium-orbit satellite, which should be taken into consideration for the acquisition algorithm of the Luojia-1A satellite.

## Methods

As can be seen from the previous section, the Doppler frequency shift range of the low-orbit satellite is large, and the rate of change of Doppler frequency shift rate is high. Therefore, the state of art parallel code phase search acquisition algorithm is presented first. Besides, the variation of power spatial density is also large according to the law of free space propagation of signals as demonstrated above. When the satellite is too far away from the ground station, the signal is too weak to acquire, so integration is adopted to improve the gain of the signal. Therefore, several major integration strategies are presented and compared to explore the signal acquisition effects of these methods on low-orbit navigation augmentation satellites such as the Luojia-1A satellite.

### Parallel Code Phase Search Acquisition

According to the implementation of code correlation and carrier correlation, there are three acquisition search algorithms: linear search, parallel frequency search, and parallel code phase search. Parallel code phase search also called circular correlation search [34],[38], can greatly reduce the computational burden and shorten the search time compared with the other two methods. All the acquisition algorithms implemented in this research are based on this search algorithm. The flow chart of the search algorithm is shown in Fig. 3.

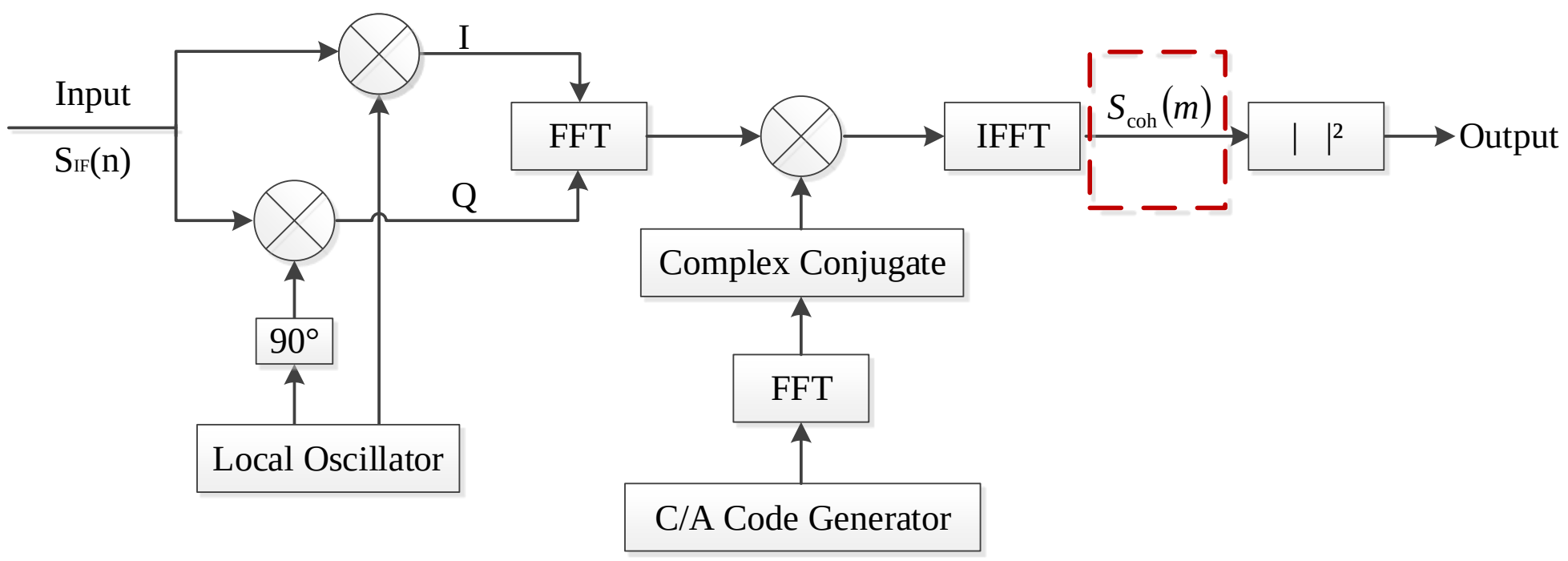

**Fig. 3.** Flow chart of parallel code phase search acquisition [37]. FFT: Fast Fourier transform; IFFT: Inverse Fast Fourier transform; I: in-phase signal; Q: quadrature signal.

The algorithm only needs to perform an iterative search on the carrier frequency without iteration on the code phase. Besides, the complex conjugate of the FFT of C/A code can be generated in advance to speed up the search process. For the convenience of operation, input data of length 1 ms, corresponding to one code length, is adopted as a processing unit. The code phase accuracy of the acquisition algorithm is related to the sampling rate of the data. The estimated code phase error of 1-millisecond coherent integration does not exceed half a sampling interval, and for the data in this work, it does not exceed one-tenth of a chip length. The frequency search bandwidth is 500 Hz, and its estimation error is less than 250 Hz. The Doppler shift accuracy of the acquisition algorithm is related to the integration time of the acquisition algorithm, and the frequency search bandwidth is inversely proportional to the integration time. As the integration time $T$ increases, the frequency search bandwidth $f_{bin}$ will shrink, which can be simplified as $f_{bin} = 500/T$.

For normal signal acquisition, the IFFT output of such a processing unit, the position at which the peak is obtained after modulo is the input signal code phase. However, for the acquisition of weak signals, it is difficult to complete the acquisition process by using only one processing unit, and it takes several processing units to complete the acquisition process. The dashed box in the figure represents the operation for weak signal acquisition, and the IFFT output of each processing unit is adopted as an input to the operation. $Soch(m)$ is the $mth$ output or result of the processing unit. The operation for the acquisition of weak signals within the dashed box is described in detail below.

## Strategies for Weak Signal Acquisition

For unaided weak signal acquisition, the receiver sensitivity can be increased by extending the integration duration [38], [39]. However, due to the bit transition and the high Doppler frequency shift rate of LEO, the integration duration cannot be extended indefinitely, and the integration acquisition process needs to be accomplished in as short a time as possible.

### *Non-coherent Integration*

Non-coherent integration is a method of increasing the signal-to-noise ratio gain by using the results of several successive processing units described in the previous section. The observation data for a long period is divided into several processing units and processed separately; then the absolute values of the processing results are accumulated as the detection value. The non-coherent operation is described as the following expression:

$$Oper_{ncoh}\left(T_{unit},T_{ncoh}\right)=\sum_{m=1}^{M}\left|S_{coh}(m)\right| \tag{6}$$

where $M$ denotes the number of processing units, which is determined by the data length of a single processing unit $T_{unit}$ and the length of the entire integration $T_{ncoh}$. Since the non-coherent integration accumulates the absolute value of the result of each processing unit, it is less affected by the bit transition. Since the incoherent integration is little affected by the bit transition, the theoretical integration duration is not limited, but the non-coherent integration has a square loss, suppressing the signal-to-noise ratio gain of the weak signal [40].

### *Coherent Integration*

The processing unit described above is a process of coherent integration with an integration duration of 1 ms. For longer coherent integration, a description similar to the non-coherent integration operation is as follows:

$$Oper_{coh}\left(T_{unit},T_{coh}\right)=\left|\sum_{m=1}^{M}S_{coh}(m)\right| \tag{7}$$

where the meaning of symbols in this expression is the same as the symbols in (6). The long-term observation data is divided into several processing units, which are processed separately; then the processing result is accumulated, and finally, the absolute value of the accumulated value is adopted as the detection value. However, unlike non-coherent integration, coherent integration acquisition may fail due to the bit transition. Therefore, variants of some coherent integration acquisition algorithms have emerged to eliminate or reduce the effects of bit transition on coherent integration. Two improved algorithms based on the coherent integration acquisition algorithm are described: the alternate half-bit method and the pre-guess test method.

As can be seen from the signal model introduced in the second section, the time interval at which bit transition occurs is a multiple of 20 ms. For a signal of 20 ms in succession, if a bit transition occurs in the first 10 ms, it is unlikely to occur in the last 10 ms. The alternate half-bit method is based on the above idea. First, the data needs to be divided into several blocks at intervals of 10 ms. As shown in Fig. 4 below, the entire data is divided into 2n blocks. Then coherent integration is performed for each data block, as described in (8).

$$y_{block}(i)=Oper_{coh}(1\text{ms},10\text{ms})\left(S_{block}(i)\right) \tag{8}$$

where the meaning of symbols in this expression is the same as the symbols in (6); $S_{block}(i)$ is the block divided as Fig. 4, and $i=1,2,\ldots 2n-1,2n$ is the index of the block; $y_{block}(i)$ is the coherent integration of each block. For the above coherent integration results, according to the odd and even blocks, the non-coherent integration is performed separately, expressed as follows:

$$\begin{cases} y_{ncoh-odd}=Oper_{ncoh}(1\text{ms},10\text{ms})(\text{odd blocks}) \\ y_{ncoh-even}=Oper_{ncoh}(1\text{ms},10\text{ms})(\text{even blocks}) \end{cases} \tag{9}$$

The non-coherent integration results $y_{ncoh-odd}$ are compared with $y_{ncoh-even}$, where large results are free of bit transition and are adopted as the final detection value. This method can avoid the effect of bit transition, but the data utilization is only 50%, and the noise power is amplified in the non-coherent process.

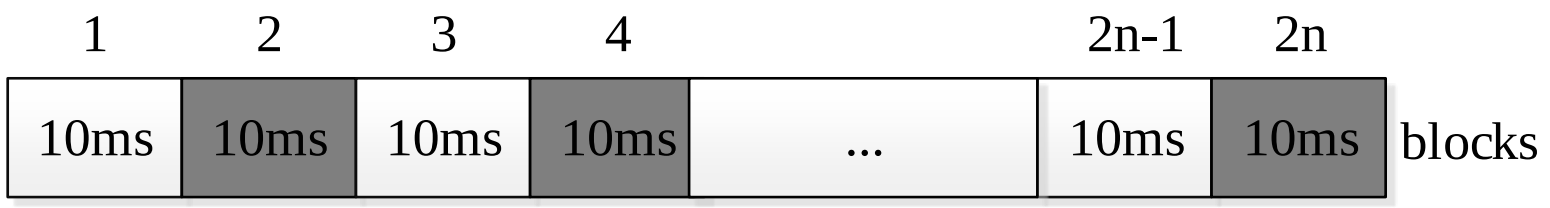

**Fig. 4.** Block division of the alternate half-bit method

The pre-guess test method is to detect and handle the problem of bit transition unit by unit. According to the existence and absence of a bit transition in the current processing unit, two coherent accumulating results are calculated, and the results are compared and determined for the following processing. The entire operation process is as follows:

$$Oper_{guess}\left(T_{unit},T_{coh}\right)=\left|\sum_{m=1}^{M} sign(m)S_{coh}(m)\right| \tag{10}$$

where the meaning of symbols in this expression is the same as the symbols in (6); $sign(m)$ is the symbol, take $+1$ or $-1$, which can be determined by

$$sign(m)=\begin{cases}+1, \left|\sum_{n=1}^{m-1} sign(n)S_{coh}(n)+S_{coh}(m)\right| > \left|\sum_{n=1}^{m-1} sign(n)S_{coh}(n)-S_{coh}(m)\right| \\ -1, \text{otherwise}\end{cases} \tag{11}$$

Since the problem of bit transition is considered in each processing unit, the influence of bit transition can be effectively eliminated incoherent integration. However, each processing unit adds additional accumulation and comparison operations, thus increasing the computation burden.

*Differential Coherent Integration*

There is also a technique called differential coherent integration that can take into account the advantages and disadvantages of the two methods mentioned above. In differential coherent integration, the processing results of adjacent processing units are conjugate multiplied, and the conjugate multiplication result is used as a new integral unit of coherent integration[39]-[42]. The operation is as follows:

$$Oper_{diff}\left(T_{unit},T_{coh}\right)=\left|\sum_{m=1}^{M} S_{coh}^{*}(m-1)S_{coh}(m)\right| \tag{12}$$

where the meaning of symbols in this expression is the same as the symbols in (6). $S_{coh}^{*}(m-1)$ is the conjugate of $S_{coh}(m)$. This method, on the one hand, can reduce the square loss of non-coherent integration; on the other hand, the effect of the bit transition of traditional coherent integration can be mitigated.

Detection Indicators

In the previous section, several integration strategies based on parallel code phase search are introduced. To compare the effects of these integration processing strategies, appropriate detection indicators are selected in this section. To describe these detection indices uniformly, the correlation values of all searched grid points are given, and the expressions are as follows:

$$R\left(\delta f_i, \delta t_j\right) \tag{13}$$

where $\delta f_i$ denotes the $ith$ Doppler shift in the frequency search range; $\delta t_j$ denotes the $jth$ code phase delay in the code phase search range.

*Maximum-to-Second-Maximum Ratio (MTSMR)*

The ratio value between the maximum correlation value and the second maximum correlation value (MTSMR) is a widely used detection index in GNSS signal acquisition [43,44]. The definition is as follows:

$$\gamma_{ratio} = \frac{R_{max}}{R_{sub}} \tag{14}$$

where $R_{max}$ and $R_{sub}$ represent the maximum correlation value and the sub-maximum correlation value, respectively. The maximum correlation value $R_{max}$ can be achieved by the following formula:

$$R_{max} = R\left(\delta f_{i_{ma}}, \delta t_{j_{ma}}\right) = max\left(R\left(\delta f_i, \delta t_j\right)\right) \tag{15}$$

where max is the maximum mathematical operator; $\delta f_{i_{ma}}$ and $\delta t_{i_{ma}}$ denote the Doppler frequency shift and code phase delay at the maximum correlation value, $i_{max}$ and $j_{max}$ are the corresponding indexes of the searching ranges. The second maximum correlation value $R_{sub}$ can be achieved as follows:

$$R_{sub} = \max\left(R\left(\delta f_{i_{ma}}, \delta t_j\right)\right), j \notin \left[j_{max} - l_{spc}, j_{max} + l_{spc}\right] \tag{16}$$

where $l_{spc}$ denotes the number of samples per code chip.

*Maximum-to-Mean Ratio (MTMR)*

The maximum-to-mean ratio is defined as follows:

$$\gamma_{mm} = \frac{R_{max}}{R_{mm}} \tag{17}$$

where $R_{mm}$ is the mean of the correlation values, excluding the peak correlation value as well as the nearby correlation values, which can be achieved by the formula as follows:

$$R_{mm} = mean\left(R\left(\delta f_i, \delta t_j\right)\right), i \notin \left[i_{\max} - 1, i_{\max} + 1\right] \text{and } j \notin \left[j_{\max} - l_{spc}, j_{\max} + l_{spc}\right] \tag{18}$$

In this formula, the 'mean' is the mean mathematical operator, It can be seen from the definition of maximum-to-mean ratio that it reflects the relative level between signal and noise from a statistical point of view. Under the condition of the fixed signal system and integration length, different integration strategies will also affect the acquisition results. The effect of these integration strategies in LEO satellite signal acquisition is presented below.

## Experiments and results

To study the use of LEO for navigation enhancement, a series of experiments were conducted to collect the Luojia- 1A satellite signal. At present, there is only one satellite of the Luojia series, namely the Luojia-1A test satellite. Due to the short transit time of the low-orbit satellite, all instruments were deployed ahead of time to wait for satellite transit.

### Test Bench for Luojia-1A Signal Sampling

In this work, a universal software radio peripheral (USRP) based test platform is designed for signal sampling and recording. USRP is an Ettus Research product, which is a low-cost, flexible, and tunable transceiver for designing, prototyping, and deploying radio communication systems. The USRP is designed to make ordinary computers work like high-bandwidth software radios. In the presented test bench, we use USRP X310, which has integrated a motherboard and two daughter boards. The USRP motherboard is responsible for clock generation and synchronization, digital-analog signal interface, host processor interface, and power management, while the USRP daughter board is used for up/down conversion, analog filtering, and other analog signals conditioning operations [45].

On July 26, 2019, data collected for about 8 minutes was stored as a file, which serves as a data source for algorithm validation described in the previous section. In this way, the process of algorithm verification is greatly simplified. The experimental configuration is shown in Fig. 5. The location of data collection is located at a ground station in Wuhan City, as shown in subfigure (a) in Fig. 5. The antenna used in the experiment is active, so an uBlox is connected to the splitter to power the antenna. The Clock Distribution Accessory 2990 (CDA-2990), also designed by Ettus Research, is an eight-channel clock distribution accessory for synchronizing multiple software radio systems and providing 1 pulse per second (PPS) time reference signals. The GPS Ant Input Interface of CDA-2990 is connected to the splitter. The frequency outputs are connected to different USRPs for devices synchronizing, and PPS outputs are connected to USRPs for timing. USRP interacts with the host computer through USRP Hardware Driver (UHD).

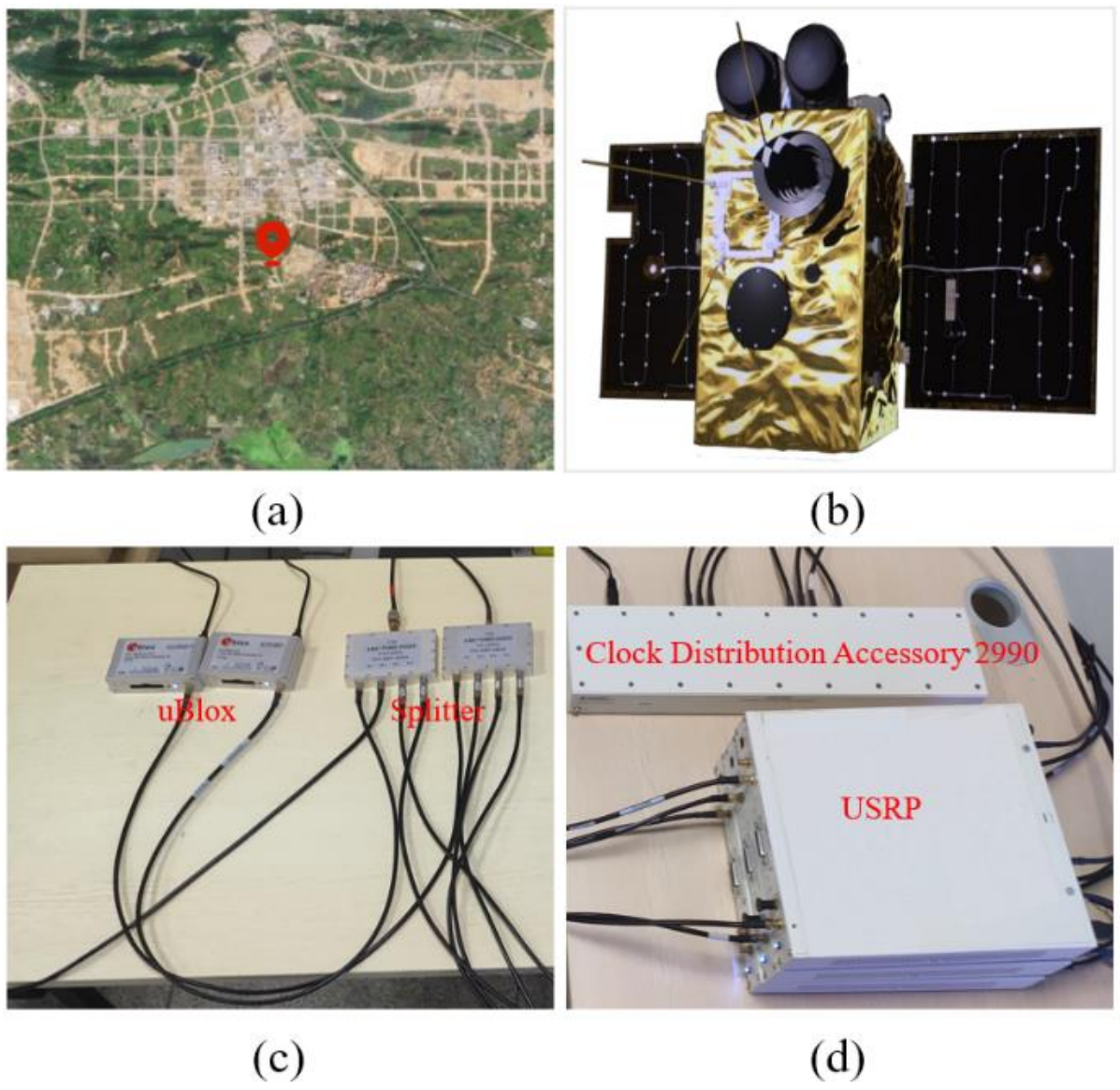

**Fig. 5.** Experimental configuration: (a) Ground station; (b) Luojia-1A satellite prototype; (c) uBlox and splitter; (d) USRP and CDA.

## Results of Process Unit

To get the performance of the integration unit with an acquisition code length - that is, the acquisition effect of the processing unit mentioned above, the collected data is coherently integrated every 1 s, and the integration length is 1 ms. The acquisition results are shown in Fig. 6.

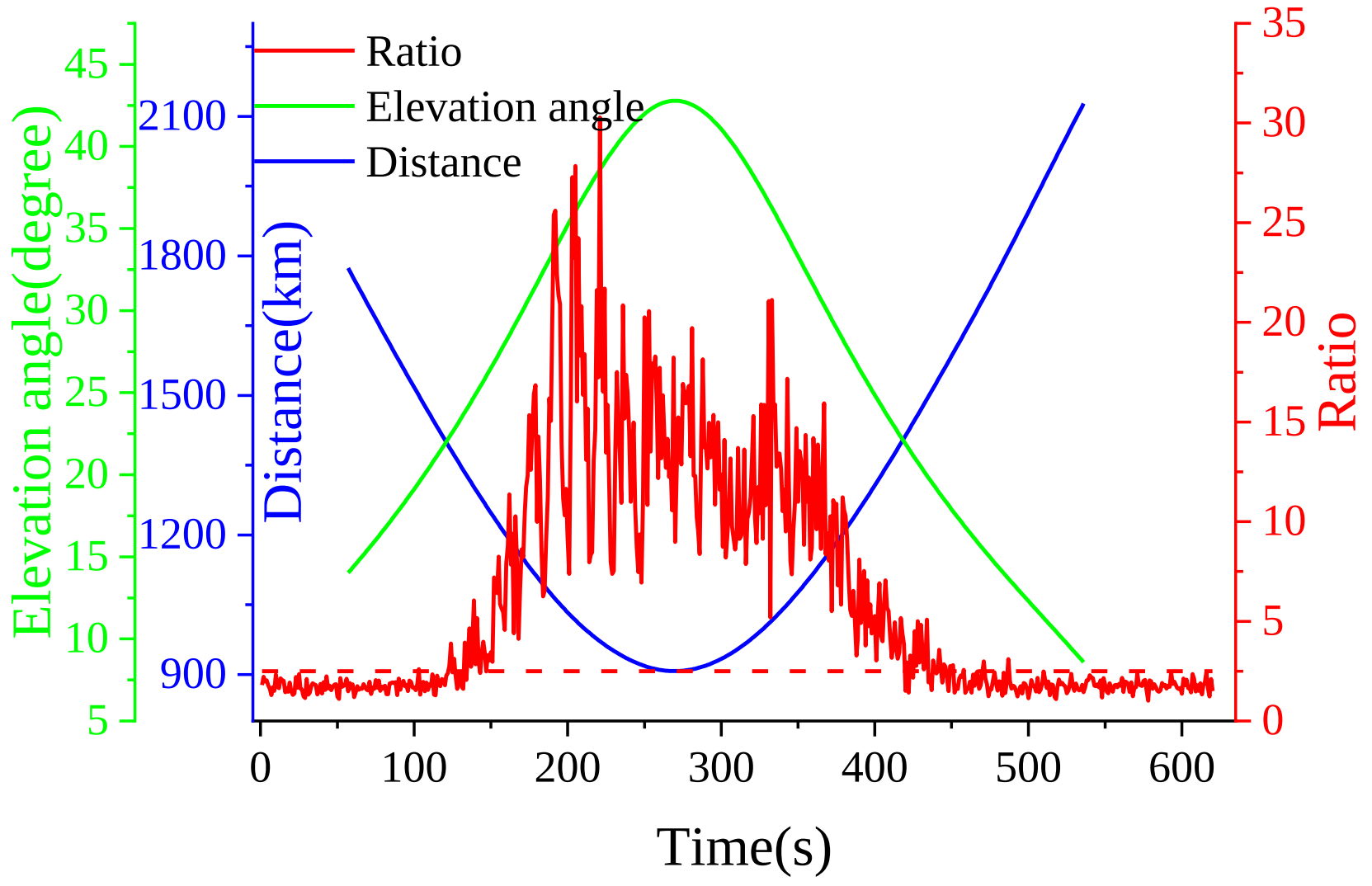

**Fig. 6.** Acquisition results of 1 ms coherent integration. Red: the maximum-to-second-maximum ratio of each second; Green: the elevation angle of each second; Blue: the distance between the ground station and the satellite at each second

In Fig. 6, the red, green, and blue lines represent the maximum-to-second-maximum ratio values, elevation angles, and distances between the ground station and the satellite, respectively. All these values are calculated at every second. The red dashed line represents the maximum-to-second-maximum ratio threshold, which is used to judge whether the acquisition is successful or not. Only signal acquisitions with a ratio value greater than this threshold are considered successful. In this paper, the maximum-to-second-maximum ratio threshold is 2.5. From the above results, it can be concluded that with the increase of the distance between the station and the satellite, the elevation angle decreases, the ratio value decreases, and the acquisition results deteriorate. The total data time is about 10 minutes, but the time interval to ensure successful acquisition is from 135 seconds to 419 seconds, a total of 285 seconds, less than 5 minutes.

## Results of 5 ms Integration

Because the LEO transit time is very short, it is of great significance to expand the successful acquisition time range of the LEO navigation augmentation signal. To study the integration strategies for expanding the successful acquisition time range, different integration strategies are used for 5 ms integration. The ratio values acquired by different integration strategies and the acquired Doppler shift results are shown in Fig. 7.

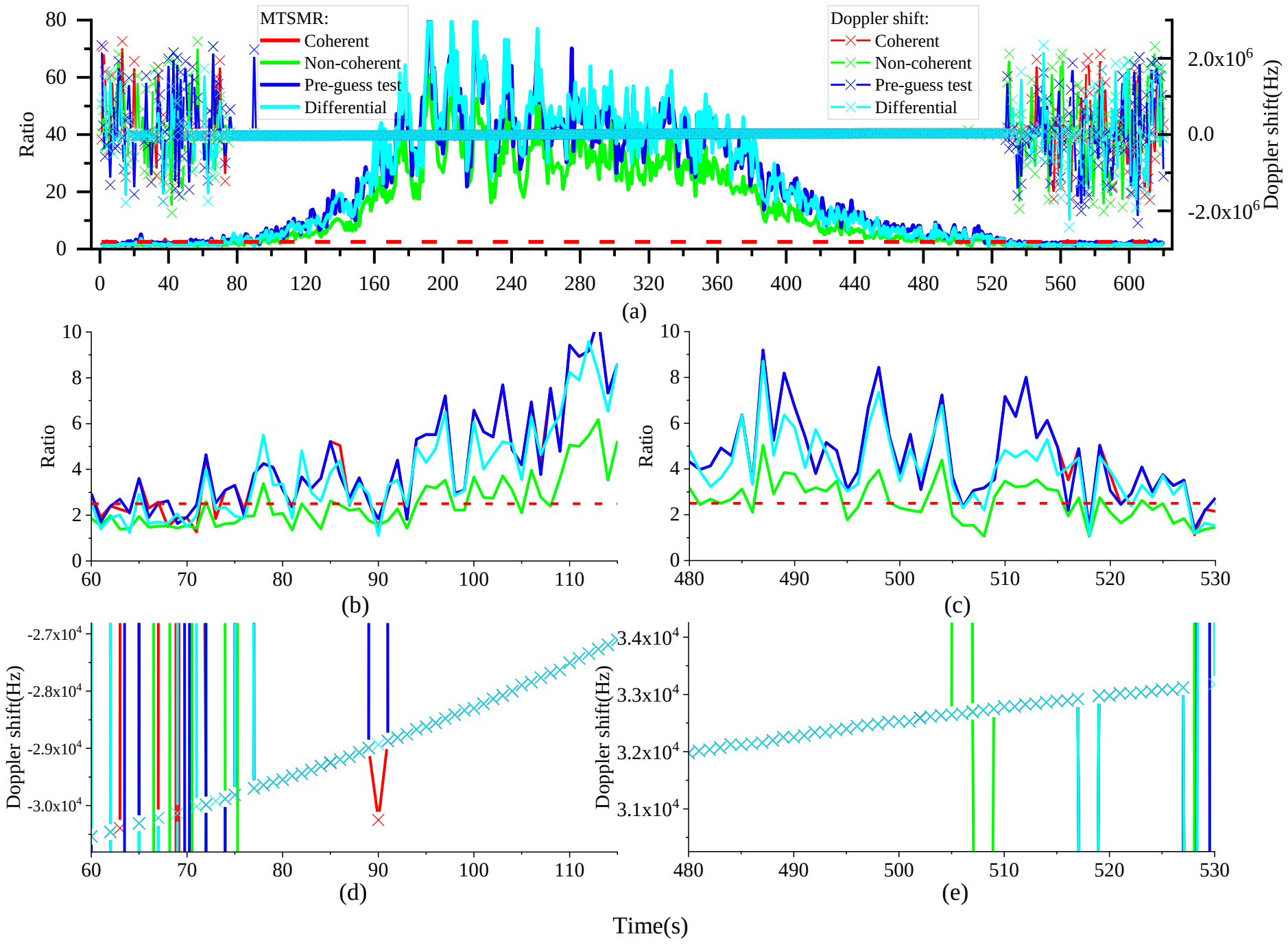

**Fig. 7.** Acquisition results of 5 ms of different integration strategies. Red: the coherent integration; Green: the non-coherent integration; Blue: the pre-guess test integration; Cyan: the differential coherent integration.

In Fig. 7, the red, green, blue, and cyan lines represent the acquisition results of coherent integration, non-coherent integration, pre-guess test integration, and differential coherent integration, respectively. All these values are calculated at every second. The line in the figure indicates the acquisition ratio value, as shown in the (b) and (c) subgraphs in the figure, and the left Y-axis of the (a) subgraph. The line combined symbol in the figure indicates the acquisition Doppler shift, as shown in the (d), and (e) subgraphs in the figure, and the right Y-axis of the (a) subgraph. The (a) subgraph is an overview of the acquisition results of various integration strategies, with the left Y-axis representing the acquisition ratio and the right Y-axis representing the acquisition Doppler shift. To more finely present the available duration of acquisition under various integration strategies, the ratio values of transition time from available to unavailable are presented in (b) and (c) subgraphs, and the corresponding Doppler shifts are presented in (d) and (e) subgraphs.

From subgraphs (b) and (c), it can be concluded that the overall successful acquisition interval is from 94 seconds to 486 seconds, lasting 393 seconds. Compared with the result of the coherent integration of 1ms, it has been greatly improved. By comparing subgraphs (b) and (d), (c), and (e), it can be found that some detection values are less than the ratio threshold, but the acquisition Doppler shift remains continuous. There is a misjudgment by setting the ratio threshold. Taking the results of incoherent integration as an example, although there are many ratio values below 2.5 between 91 seconds and 108 seconds, 486 seconds, and 505 seconds, the acquisition Doppler shift remains continuous and can be acquired correctly. The selection and setting of the threshold are discussed in the following section. It can be seen from the results of the acquired Doppler shift in subgraph (d) that at 90 seconds, all methods except the differential coherence method fail. It can be seen from the results of the acquired Doppler shift in subgraph (e) that at 505 seconds, all methods except the incoherent method are successfully acquired. It can be seen from the Doppler shift that the total available time is between 77 and 527 seconds, lasting 450 seconds.

## Results of 20 ms Integration

The ratio and Doppler shift results of 20 ms integration are shown in Fig. 8. The integration results of the alternate half-bit method are also shown in the figure. The display details in Fig. 8 are similar to those in Fig. 7, except that the newly added black element represents the result of the alternate half-bit method. As can be seen from the above figure, the period during which the signal can be successfully acquired by the alternate half-bit method is from 45 seconds to 537 seconds. The alternate half-bit method acquired period coincides with the differential coherent acquired period. Due to the misjudgment of the bit transition, the performance of the pre-guess test is the worst. Similar to the 5 ms integration results, although some of the detected values are lower than the threshold 2.5, the signal is successfully acquired, which is more obvious in the non-coherent integration.

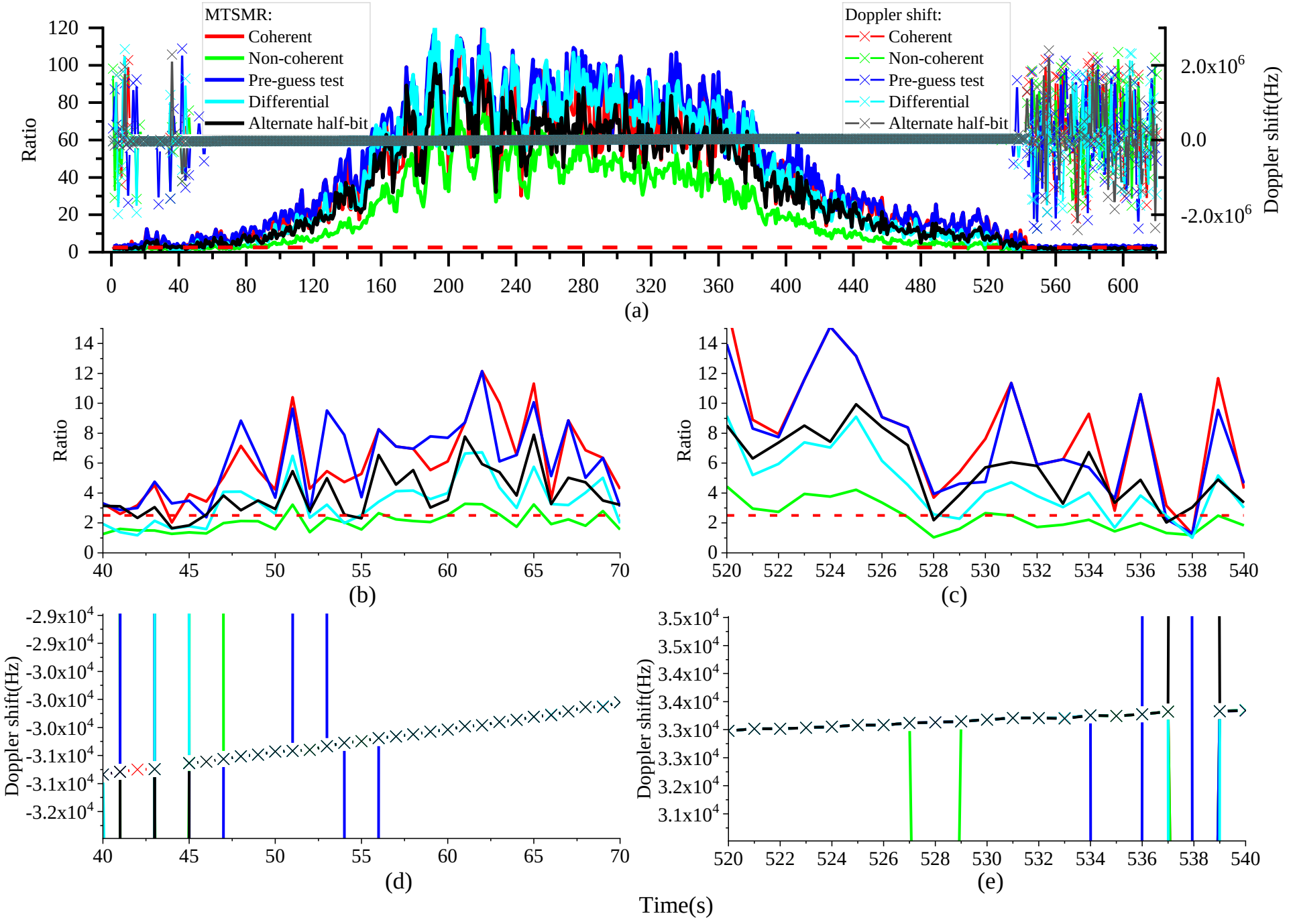

**Fig. 8.** Acquisition results of 20ms of different integration strategies. Red: the coherent integration; Green: the non-coherent integration; Blue: the pre-guess test integration; Cyan: the differential coherent integration; Black: the alternate half-bit integration.

Results of Other Integration Duration

For different integration durations, the successful acquisition time length and range of different integration algorithms are summarized, as shown in Table. 1 and Fig. 9. The successful acquisition time range is calculated based on the continuity of the Doppler shift. It can be seen from these results that the integration can effectively expand the range of successful acquisition, thus increasing the available time of the signal. However, due to the influence of noise, the time range of successful acquisition will not expand infinitely with the increase of integration duration.

**Table 1** Successful acquisition time length of different integration algorithms for different integration time lengths.

| integration strategies | 1ms | 2ms | 5ms | 10ms | 15ms | 20ms | 30ms | 40ms |
|---|---|---|---|---|---|---|---|---|
| Coherent | 327s | 393s | 460s | 510s | 522s | 522s | 509s | 498s |
| Non-coherent | | 390s | 445s | 485s | 495s | 505s | 504s | 493s |
| Pre-guess test | | 393s | 454s | 482s | 495s | 502s | 505s | 498s |
| Differential | | 416s | 465s | 501s | 513s | 514s | 506s | 494s |
| Alternating half bit | | | | | | 514s | | 494s |

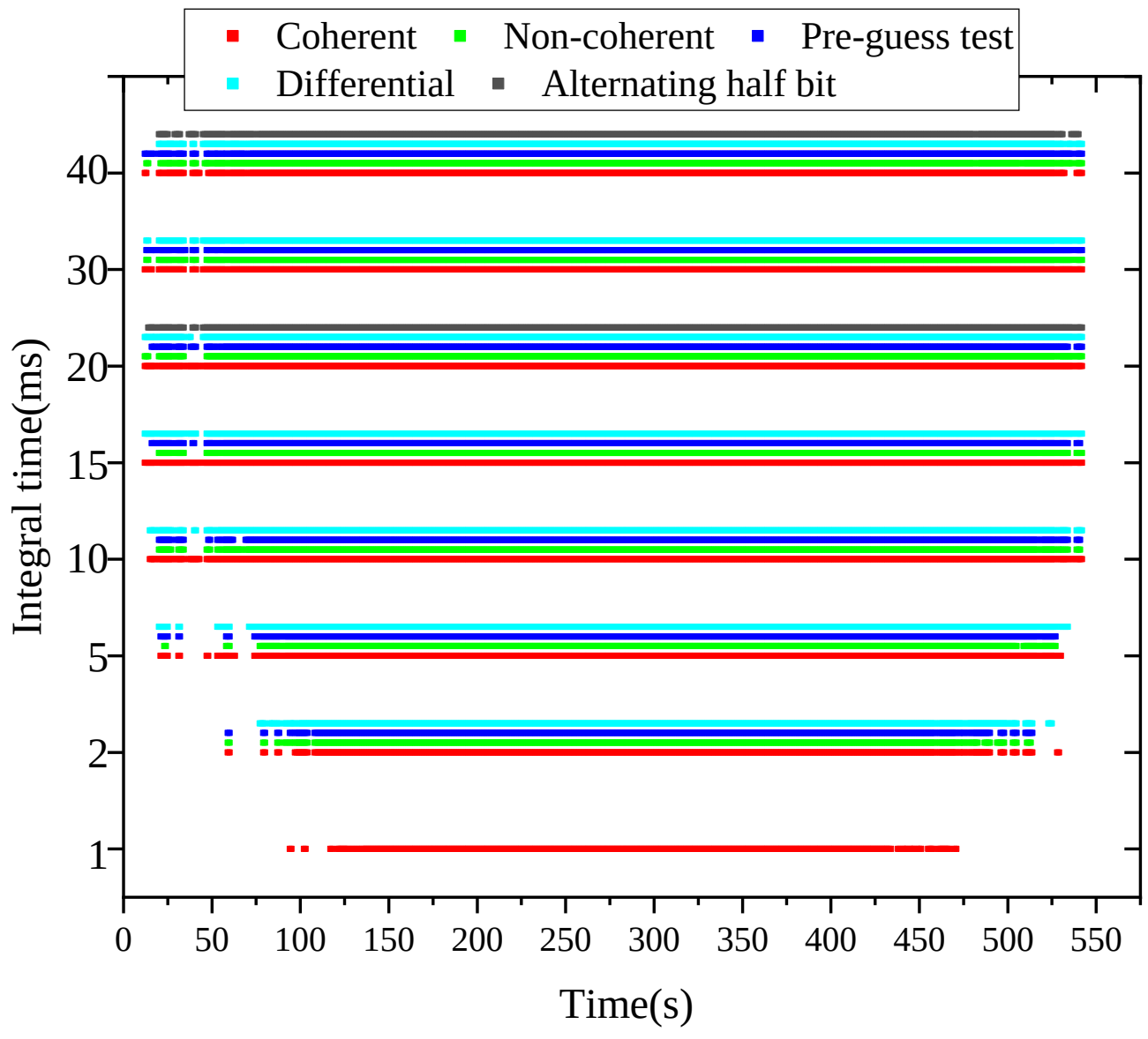

**Fig. 9.** Successful acquisition time range of different integration algorithms for different integration duration lengths

## Discussion

In the above results, the detection index is MTSMR, and the threshold of the detection index is an empirical value of 2.5. However, it can be found that the MTSMR output values of the different integration algorithms are significantly different, with the non-coherent MTSMR values being significantly smaller than the other integration strategies. At the same time, it is found that many MTSMR detection values are less than the threshold value, but the obtained Doppler shift remains continuous, that is, successfully obtained. Therefore, it is necessary to explore the reasonable setting of those indicators' thresholds and whether those thresholds are related to the integration duration.

### Thresholds of Detection Indicators

As shown in the previous section, the range of MTSMR detection value varies with different integration algorithms, such as the MTSMR detection value of non-coherent integration is significantly smaller than that of other integration algorithms. Based on the continuity of the Doppler shift, the probability of false alarm ( $P_f$ ) using different MTSMR thresholds under different integration algorithms of 5 ms is given, as shown in Fig. 10.

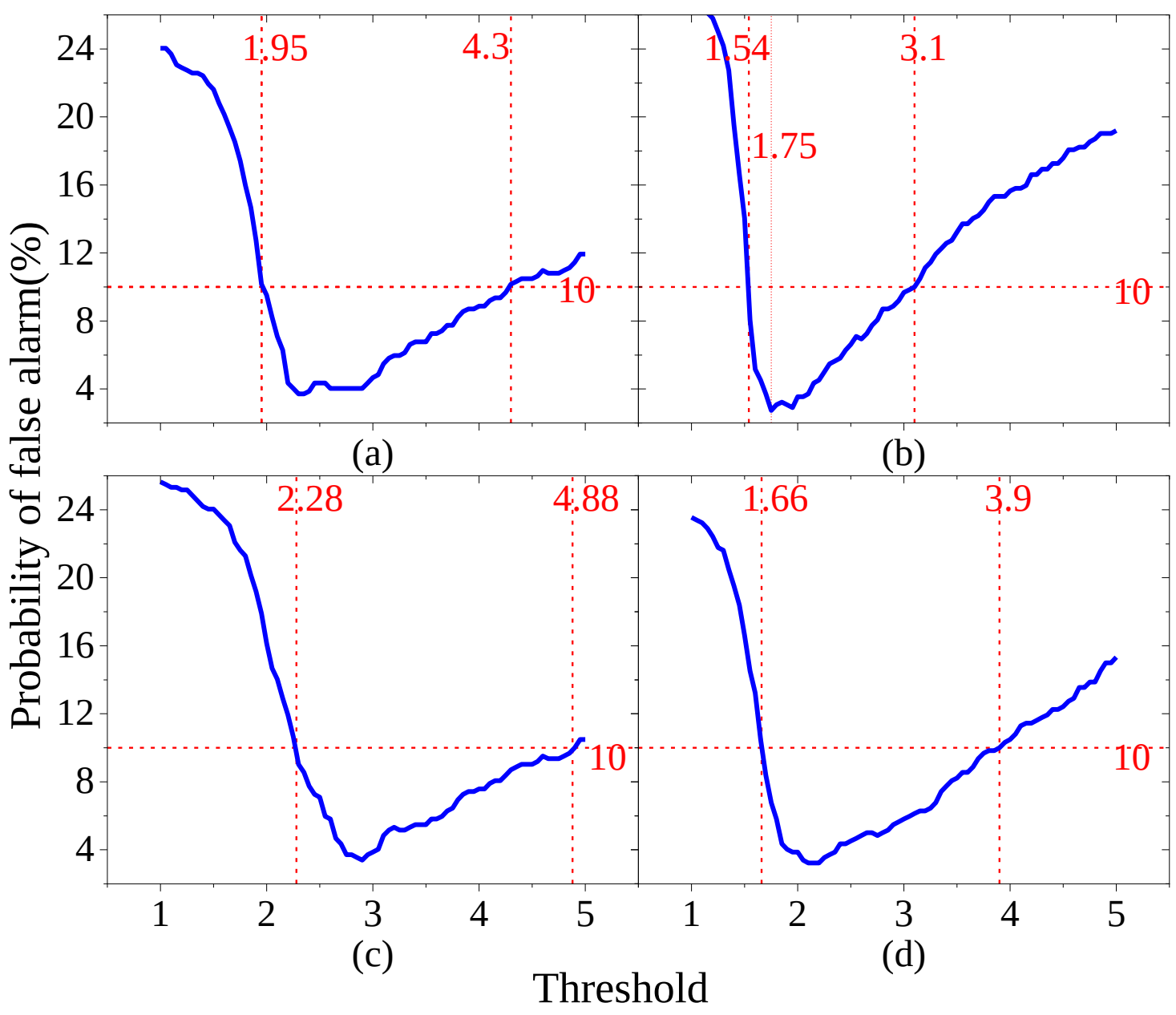

**Fig. 10.** Probability of false alarm using different MTSMR thresholds under different integration algorithms of 5 ms. (a): the coherent integration; (b): the non-coherent integration; (c): the pre-guess test integration; (d): the differential coherent integration.

In Fig. 10, the horizontal axis represents the threshold of MTSMR and the vertical axis represents the $P_f$. The horizontal red dot lines represent 10% $P_f$, and vertical red dot lines represent the critical threshold for obtaining 10% $P_f$. Taking the noncoherent integration as an example, when the detection threshold of MTSMR is between 1.54 and 3.1, the non-coherent integration of 5 ms can achieve the $P_f$ less than 10%, that is, the probability of detection ($P_d$) is more than 90%. When the threshold is 1.75, the lowest $P_f$ value can be obtained: 2.7419%, that is, the $P_d$ reaches the maximum value: 97.2581%. If the threshold is set too large or too small, the $P_f$ will increase. When the threshold setting is too large, it is easy to detect a successful acquisition as a failed acquisition. When the threshold setting is too small, it is easy to detect a failed acquisition as a successful acquisition. To further discuss the relationship between the MTSMR threshold and the integration duration, the relationship between the integration duration and thresholds of less than 10% $P_f$ is given, as shown in Fig. 11.

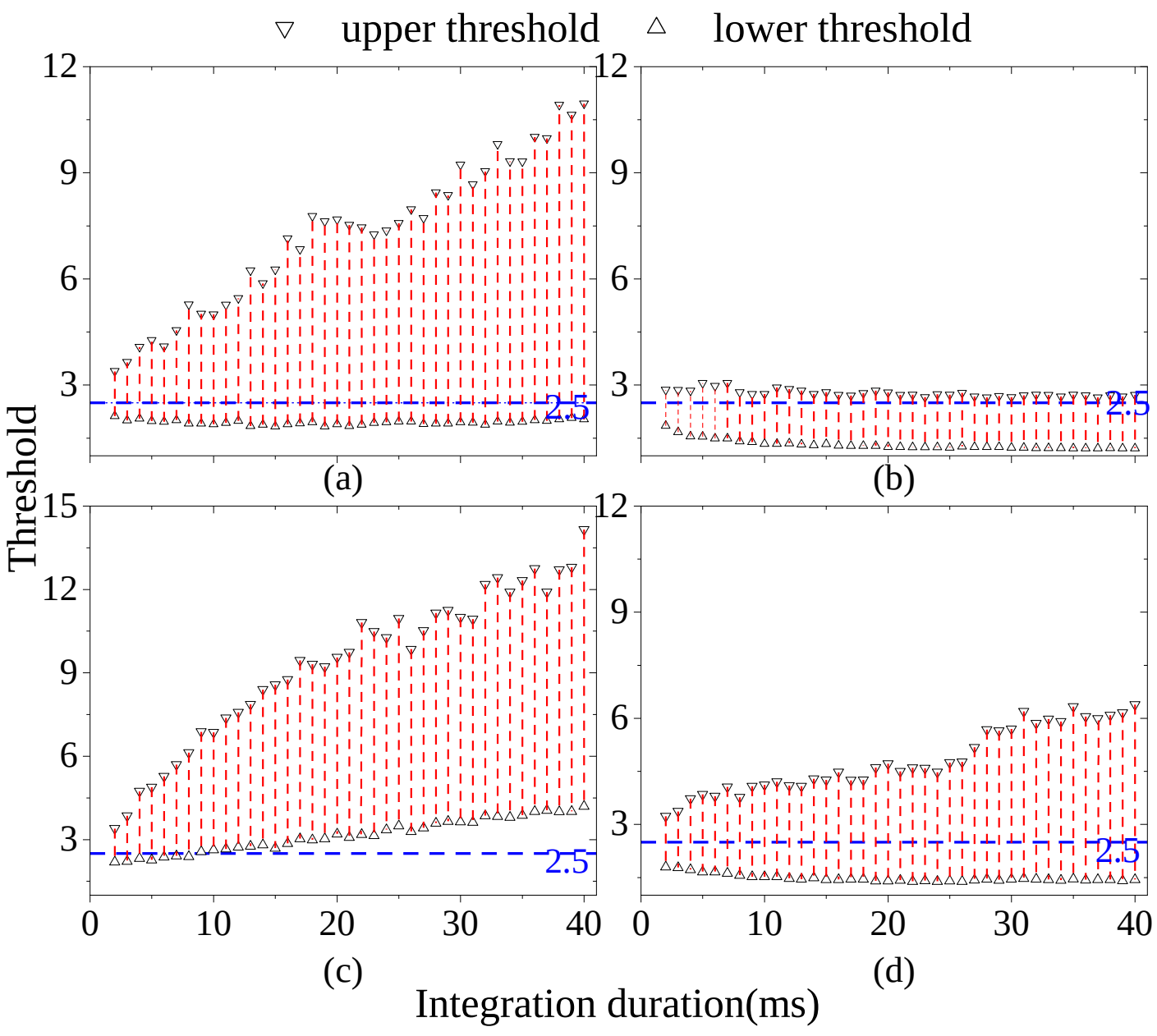

**Fig. 11.** The upper and lower MTSMR thresholds of less than 10% Pf for various integration duration. (a): the coherent integration; (b): the non-coherent integration; (c): the pre-guess test integration; (d): the differential coherent integration.

In Fig. 11, the horizontal axis denotes the integration duration and the vertical axis represents the threshold of MTSMR. For coherent integration, non-coherent integration, and differential integration, the $P_f$ is less than 10% in the integration duration of 2-20 ms when the MTSMR threshold value is 2.5. For the pre-guess test integration, when the threshold is selected to be 2.5, the $P_f$ is greater than 10% when the integration duration is greater than 9 ms. Under the premise that the $P_f$ is less than 10%, with the prolongation of integration duration, the range of optional threshold of coherent integration increases gradually, the range of optional threshold of noncoherent integration is smaller and relatively stable, the range of optional threshold of pre-guess test gradually expands and tends to move upward, and the range of differential coherence threshold is relatively stable. For the other detection indicator MTMR mentioned above, the relationship between the integration duration and thresholds of less than 10% $P_f$ is also given, as shown in Fig.12.

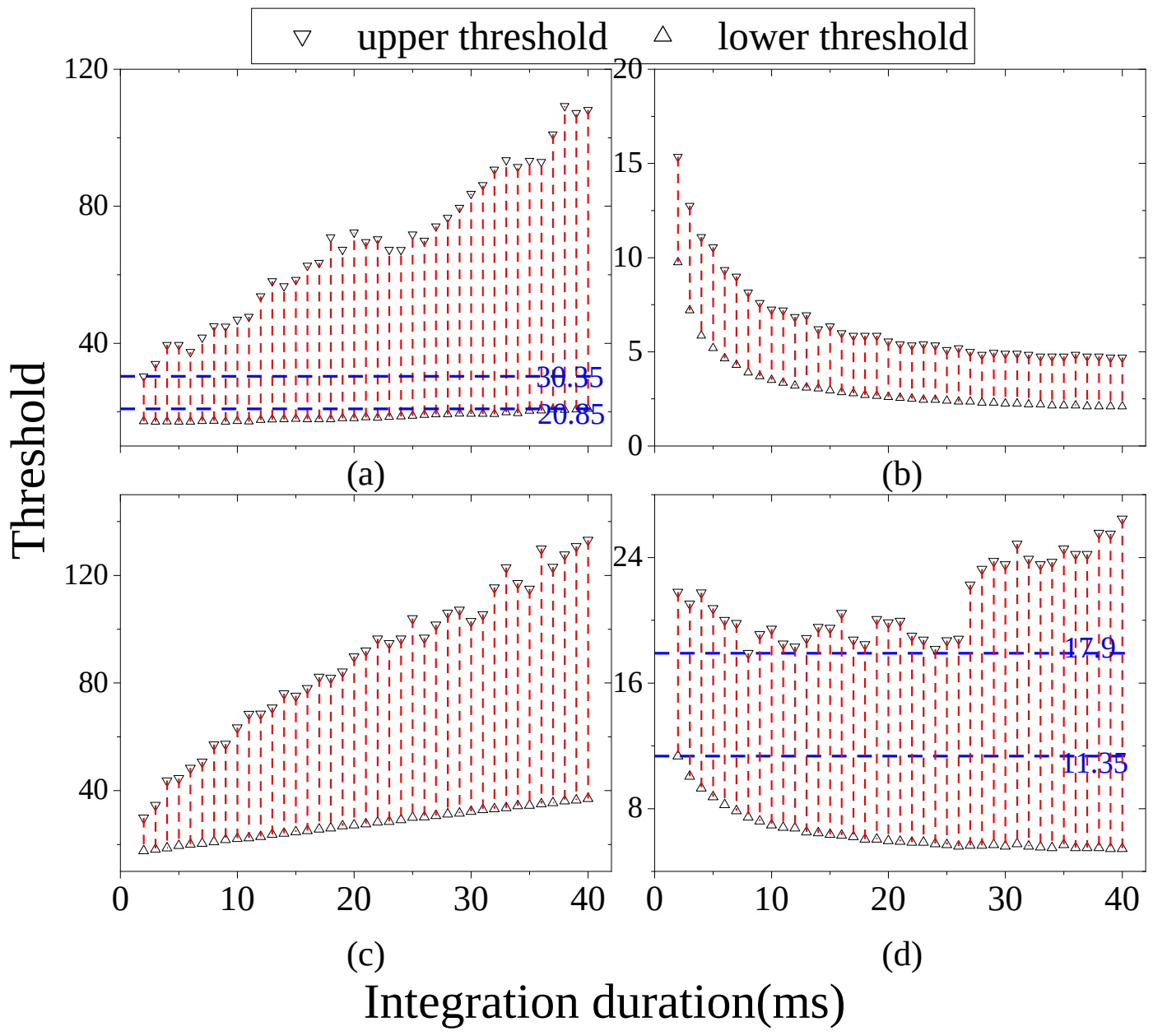

**Fig. 12.** The upper and lower MTMR thresholds of less than 10% Pf for various integration duration. (a): the coherent integration; (b): the non-coherent integration; (c): the pre-guess test integration; (d): the differential coherent integration.

As can be seen from the figure, the obtained MTMR threshold range is significantly different in magnitude for different integration algorithms, so it is difficult to use a global value as the MTMR detection threshold for all integration algorithms. The threshold between two dot lines represents the intersection of the optional threshold ranges of different integration duration between 2 and 20 ms. Under the premise that the $P_f$ is lower than 10%, the lower limit of the optional threshold of coherent integration does not change significantly with the increase of the integration duration and the overall range of the optional threshold increases with the upper limit of the threshold. The upper and lower limits of the optional thresholds of noncoherent integration gradually decrease, and the fluctuation is large. It is difficult to use the same threshold for noncoherent integration to obtain less than 10% $P_f$ of different integration duration. The upper and lower limits of the optional thresholds of the pre-guess test method increase gradually, but the threshold range intersection of different integration duration is smaller. The lower limit of differential coherence optional threshold decreases gradually and tends to be stable, and the range of optional threshold increases.

Integration Duration

To study the relationship between integration duration and successful acquisition time, the acquisition time of 2-40 ms integration duration is calculated, which is shown in Fig. 13.

In general, with the increase of integration duration, the successful acquisition time first increases and then decreases, and the successful acquisition time will not increase indefinitely with the integration duration. When the integration duration is more than 20 ms, the length of successful acquisition time decreases. The effect of coherent integration and differential integration is better than the other two integration methods. When the integration duration is less than 9 ms, the effect of differential integration is better than that of coherent integration, and the effect of coherent integration is slightly better than the coherent integration when the integration duration is longer than 9 ms.

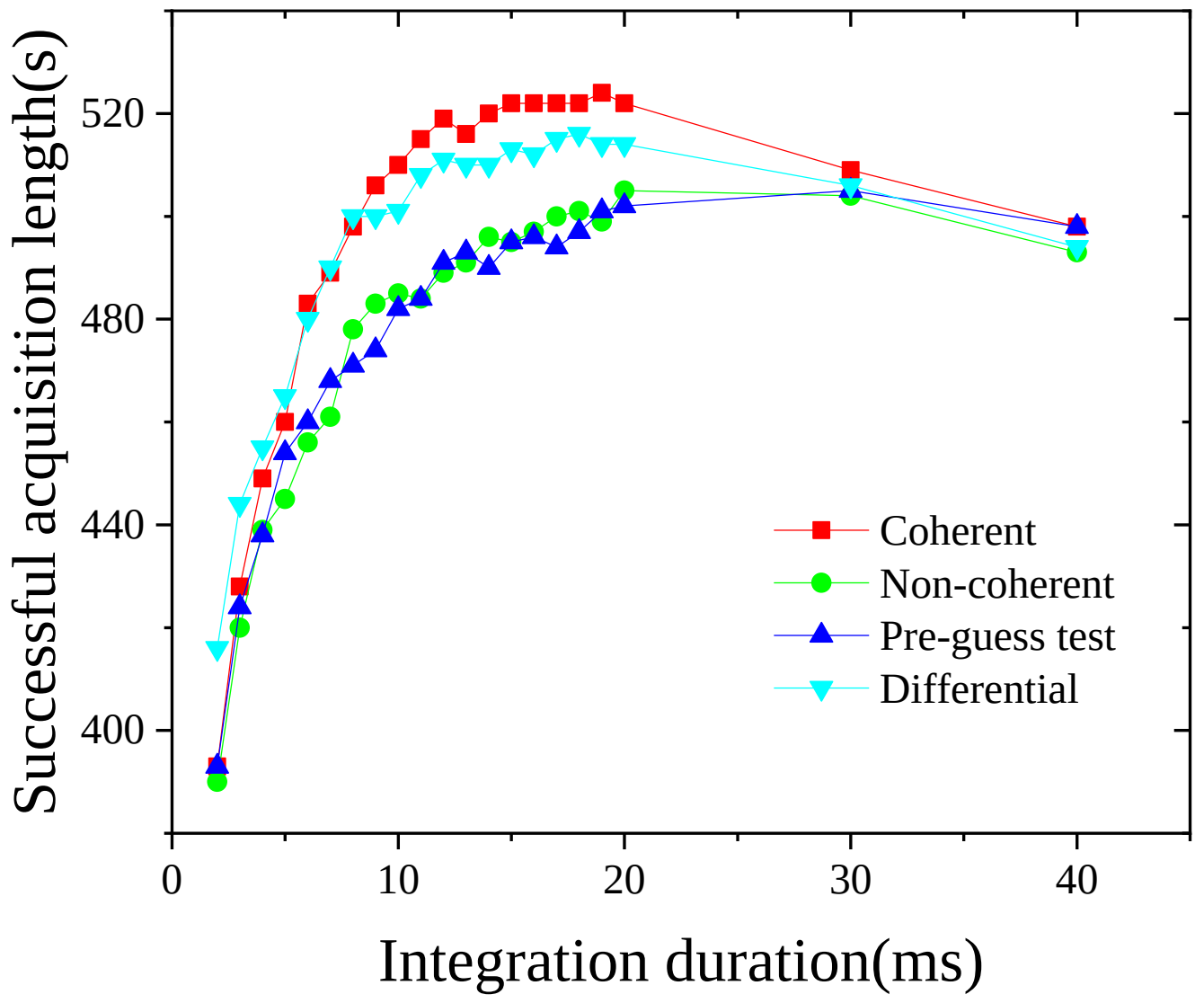

**Fig. 13.** Relationship between integration duration and successful acquisition time length

## Conclusions

This paper aims to study LEO signal acquisition and try to expand the successful acquisition time range. One of the more significant findings to emerge from this study is that the integration strategies expand the successful acquisition time range, and it will not expand indefinitely with the integration duration.

Through the study of LEO's orbit and signal characteristics, it is found that compared with medium earth orbit satellite, LEO has the characteristics of large Doppler shift and large variation in power space density. Based on the parallel code search signal acquisition algorithm, the unified symbol definitions of coherent integration, noncoherent integration, differential coherent integration, pre-guess test, and alternating half-bit algorithms are given. To verify and analyze the above acquisition algorithms, a software-defined receiver was developed and the real data was collected from the Luojia-1A satellite by using USRP. The experimental results show that the successful acquisition time range of the 1 ms integration duration is about 300 s. Integration strategies can significantly expand the successful acquisition time range, and the maximum acquisition time range can reach 522 s. However, due to the change in signal strength and the presence of bit transition, it is difficult to maintain a longer successful acquisition time range even if the integration duration is infinitely extended. Besides, the thresholds of detection indicators under different integration algorithms and various integration duration are discussed and given. For all integration algorithms except the pre-guess test, the $P_f$ is less than 10% in the integration duration of 2-40 ms when the MTSMR threshold value is adopted the empirical value 2.5 and it is difficult to use a global value as the MTMR detection threshold for all integration strategies. The trend of successful acquisition time range versus integration duration under different integration algorithms is discussed. The performance of coherent integration and differential integration is better than the other two integration algorithms.

## Disclosure statement

No potential conflict of interest was reported by the authors.

## Funding

This research was supported by National Key R&D Program of China (2018YFB0505400), the Natural Science Fund of Hubei Province with Project No. 2018CFA007.

## Data availability statement

The data that support the findings of this study are available from the State Key Laboratory of Information Engineering in Surveying, Mapping and Remote Sensing (LIESMARS), www.lmars.whu.edu.cn. The University of Wuhan, but restrictions apply to the availability of these data, which were used under license for the current study, and so are not publicly available. Data are however available from the authors upon reasonable request and with permission of LIESMARS.

## Author Biographies

**Liang Chen** was a Senior Research Scientist with the Department of Navigation and Positioning, Finnish Geodetic Institute, Finland. He is currently a Professor with the State Key Laboratory of Information Engineering in Surveying, Mapping, and Remote Sensing, Wuhan University, China. He has published over 70 scientific articles and five book chapters. His current research interests include indoor positioning, wireless positioning, sensor fusion, and location-based services. He is currently an Associate Editor of the Journal of Navigation, Navigation, and Journal of Institute of Navigation.

**Xiangchen Lu** is currently a master student at the State Key Laboratory of Information Engineering in Surveying, Mapping and Remote Sensing at Wuhan University. His main research direction is the software-defined receiver.

**Nan Shen** is currently a Ph.D. student at the State Key Laboratory of Information Engineering in Surveying, Mapping and Remote Sensing at Wuhan University. His research interests focus on precise GNSS data processing, software-defined receiver.

**Lei Wang** is currently an associate research fellow at State Key Laboratory of Information Engineering in Surveying, Mapping and Remote Sensing at Wuhan University. His research interests include GNSS precise positioning and LEO navigation augmentation system.

**Yuan Zhuang** is a professor at State Key Laboratory of Information Engineering in Surveying, Mapping and Remote Sensing, Wuhan University, China. He received the bachelor degree in information engineering from Southeast University, Nanjing, China, in 2008, the master degree in microelectronics and solid-state electronics from Southeast University, Nanjing, China, in 2011, and the Ph.D. degree in geomatics engineering from the University of Calgary, Canada, in 2015. His current research interests include multi-sensors integration, real-time location system, personal navigation system, wireless positioning, Internet of Things (IoT), and machine learning for navigation applications. To date, he has co-authored over 70 academic papers and 11 patents and has received over 10 academic awards. He is an associate editor of IEEE Access, the guest editor of the IEEE Internet of Things Journal and IEEE Access, and a reviewer of over 10 IEEE journals.

**Ye Su** is an undergraduate student from School of Remote Sensing Information Engineering, Wuhan University, Wuhan.

**Deren Li** received the M.Sc. degree in photogrammetry and remote sensing from the Wuhan Technical University of Surveying and Mapping, Wuhan University, Wuhan, China, in 1981, and the Dr.Eng. degree in photogrammetry and remote sensing from Stuttgart University, Stuttgart, Germany, in 1985. He was elected as an Academician of the Chinese Academy of Sciences, Beijing, China, in 1991, and the Chinese Academy of Engineering, Beijing, and the Euro–Asia Academy of Sciences, Beijing, in 1995. He is currently the Academic Committee Chairman of the State Key Laboratory of Information Engineering in Surveying, Mapping and Remote Sensing, Wuhan University. His research interests include spatial information science and technology, such as remote sensing, GPS and geographic information system (GIS), and their integration. He was the President of the International Society for Photogrammetry and Remote Sensing Commissions III and VI and the first President of the Asia GIS Association, from 2002 to 2006.

**Ruizhi Chen** is a professor at the State Key Laboratory of Surveying, Mapping and Remote Sensing Information Engineering, Wuhan University, and director of the laboratory. His main research interests include smartphone ubiquitous positioning and satellite navigation and positioning.